\documentclass[12pt]{article}
\usepackage[ansinew]{inputenc}
\usepackage{amsmath}
\usepackage{multirow}
\usepackage{enumerate}
\usepackage[top=2cm, bottom=2cm, right=2cm, left=2cm]{geometry}
\title{\bf Centered and non-centered variance inflation factor}
\author{Rom\'an Salmer\'on G\'omez \\
    Dpto. de M\'etodos Cuantitativos para la Econom\'ia y la Empresa \\
    Universidad de Granada \\
     \\
    Catalina Garc\'ia Garc\'ia \\
    Dpto. de M\'etodos Cuantitativos para la Econom\'ia y la Empresa \\
    Universidad de Granada \\
    \\
    Jos\'e Garc\'ia P\'erez \\
    Dpto. de Econom\'ia y Empresa \\
    Universidad de Almer\'ia
}

\begin{document}

\maketitle

\begin{abstract}
     This paper analyzes the diagnostic of near multicollinearity in a multiple linear regression from auxiliary centered regressions (with intercept) and non-centered  (without intercept). From these auxiliary regression, the centered and non-centered Variance Inflation Factors are calculated, respectively. It is also presented an expression that relate both of them.
\end{abstract}

\noindent%

\section{Introduction}
\label{sec:intro}

Considering the following multiple linear model with $n$ observations and $k$ regressors :
\begin{equation}
    \mathbf{y}_{n \times 1} = \mathbf{X}_{n \times k} \cdot \boldsymbol{\beta}_{k \times 1} + \mathbf{u}_{n \times 1},
    \label{model0}
\end{equation}
where $\mathbf{y}$ is a vector with the observations of the dependent variable, $\mathbf{X}$ is a matrix containing the observations of regressors and  $\mathbf{u}$ is a vector representing random disturbance (that is supposed to be spherical). When this model presents near multicollinearity, it is to say, when the linear relation between the regressors affects to the numerical and/or statistical analysis of the model, it is usual to transform the data (see, for example, Belsley \cite{Belsley1984}, Marquardt \cite{Marquardt1980} or, more recently, Velilla \cite{Velilla2018}).

In these cases, the first column of matrix $\mathbf{X}$ is said to be composed by ones to denote that the model contains an intercept.  Thus, $\mathbf{X} = [ \mathbf{1} \ \mathbf{X}_{2} \dots \mathbf{X}_{k}]$ where $\mathbf{1}_{n \times 1} = (1 \ 1 \dots 1)^{t}$. This model is considered to be centered. On the other hand, transformed models are considered to be non-centered, since the transformations (centering, typification or standardization) imply the elimination of the intercept in the model. Note that even after transforming the data, it is possible to recover the original model (centered) from the estimations of the transformed model (non-centered model). However, in this paper we refer to centered and non-centered model depending on the inclusion of intercept. Thus, it is considered that the model is centered if  $\mathbf{X} = [ \mathbf{1} \ \mathbf{X}_{2} \dots \mathbf{X}_{k}]$ and non-centered if $\mathbf{X} = [ \mathbf{X}_{1} \ \mathbf{X}_{2} \dots \mathbf{X}_{k}]$ being $\mathbf{X}_{j} \not= \mathbf{1}$ with $j=1,\dots,k$.\\

The main contribution of the paper is the analysis of the variance inflation factor, which is widely applied to detect multicollinearity in model (\ref{model0}) by considering that the auxiliary regression used to its calculation is centered or not. For a better understanding of this paper, could be interesting to distinguish between the following two kinds of near multicollinearity that can be found in model  (\ref{model0}), (see Marquardt and Snee \cite{MarquardtSnee}):
\begin{description}
    \item[Non essential:] Near linear relation between the intercept and at least one of the independent variables.
    \item[Essential:] Near linear relation between at least two of the independent variables (excluding the intercept).
\end{description}

The structure of the paper is as follows: Section \ref{1} presents some preliminaries and introduces the main questions that will be answered through the paper, Section \ref{sec:regression_no_centered} presents the non-centered auxiliary regressions and, finally, section \ref{sec:conc} summarizes the main conclusions.

\section{Preliminaries}\label{1}

Considering $k=3$ in (\ref{model0}), Belsley \cite{Belsley1984} used the non-centered coefficient of determination (without considering the intercept) of the regression of $\mathbf{X}_{3}$ as a function of $\mathbf{X}_{2}$ to calculate the non-centered variance inflation factor (denoted as  VIFnc) of the following regression:
$$\mathbf{y} = \beta_{1} \cdot \mathbf{1} + \beta_{2} \cdot \mathbf{X}_{2} + \beta_{3} \cdot \mathbf{X}_{3} + \mathbf{u},$$
Thus, Belsley \cite{Belsley1984} used the coefficient of determination of the following auxiliary regression:
$$\mathbf{X}_{3} = \delta \cdot \mathbf{X}_{2} + \mathbf{v},$$
taking into account the values \footnote{Variables $\mathbf{y}$, $\mathbf{X}_{2}$ and $\mathbf{X}_{3}$ were originally used by Belsley \cite{Belsley1984}. Variable $\mathbf{X}_{4}$ has been randomly generated to obtain a variable linearly independent to the rest.} displayed in Table \ref{datos.Belsley1984984}. In these data, it is intuited the existence of near non-essential multicollinearity, it is to say, relation between the intercept and at least one of the independent variables of the model.

\begin{table}
    \centering
    \begin{tabular}{ccccc}
        \hline
        $\mathbf{y}$ & $\mathbf{X}_{1}$  & $\mathbf{X}_{2}$ & $\mathbf{X}_{3}$ & $\mathbf{X}_{4}$ \\
        \hline
        2.69385 & 1 & 0.996926 & 1.00006 & 8.883976 \\
        2.69402 & 1 & 0.997091 & 0.998779 & 6.432483 \\
        2.70052 & 1 & 0.9973 & 1.00068 & -1.612356 \\
        2.68559 & 1 & 0.997813 & 1.00242 & 1.781762 \\
        2.7072 & 1 & 0.997898 & 1.00065 & 2.16682 \\
        2.6955 & 1 & 0.99814 & 1.0005 & 4.045509 \\
        2.70417 & 1 & 0.998556 & 0.999596 & 4.858077 \\
        2.69699 & 1 & 0.998737 & 1.00262 & 4.9045 \\
        2.69327 & 1 & 0.999414 & 1.00321 & 8.631162 \\
        2.68999 & 1 & 0.999678 & 1.0013 & -0.4976853 \\
        2.70003 & 1 & 0.999926 & 0.997579 & 6.828907 \\
        2.702 & 1 & 0.999995 & 0.998597 & 8.999921 \\
        2.70938 & 1 & 1.00063 & 0.995316 & 7.080689 \\
        2.70094 & 1 & 1.00095 & 0.995966 & 1.193665 \\
        2.70536 & 1 & 1.00118 & 0.997125 & 1.483312 \\
        2.70754 & 1 & 1.00177 & 0.998951 & -1.053813 \\
        2.69519 & 1 & 1.00231 & 1.00102 & -0.5860236 \\
        2.7017 & 1 & 1.00306 & 1.00186 & -1.371546 \\
        2.70451 & 1 & 1.00394 & 1.00353 & -2.445995 \\
        2.69532 & 1 & 1.00469 & 1.00021 & 5.731981 \\
        \hline
    \end{tabular}
    \caption{Data set applied by Belsley \cite{Belsley1984}} \label{datos.Belsley1984984}
\end{table}

By using the original variables applied by Belsley, the traditional VIF (from centered model, see Theil \cite{Theil1971}) provides a value equal to 1 (its minimum possible value), while the VIFnc is equal to 100032.1. However, if an additional variable is included (generated from a normal distribution with mean equal to 4 and variance equal to 16), $\mathbf{X}_{4}$, the following values are obtained for the VIF and the VIFnc of the three variables:
$$1.155364, 1.084168, 1.239559, \qquad 100453.8, 100490.6, 1.773768.$$
Thus, the VIF is not detecting the existence of essential near multicollinearity, (see Salmer\'on et al \cite{Salmeron2018}) while the VIFnc  does detect it.

However, since the calculation of VIFnc excludes the constant term, the detected relation refers to the one between $\mathbf{X}_{2}$ and $\mathbf{X}_{3}$, and not to the relation between $\mathbf{X}_{2}$ and/or $\mathbf{X}_{3}$ with the intercept. These results suggest a new definition of non-essential multicollinearity as the relation between at least two variables with little variability. Thus, the particular case when one of these variables is the intercept leads to the definition initially given by Marquardt y Snee \cite{MarquardtSnee}.

The following values are obtained for the VIF and VIFnc of the second and fourth variables, respectively:
$$1.143328, \qquad 1.765676,$$
while, for the third and fourth:
$$1.072873, \qquad 1.766323.$$
Thus, it is not detected in any case a relation between $\mathbf{X}_{2}$ or $\mathbf{X}_{3}$ with the intercept.

With these results and by following Salmer\'on et al \cite{Salmeron2018}, it can be concluded that the VIF only detects the near essential multicollinearity and the VIFnc only detects the non essential near multicollinearity. It will be interesting to analyze if the VIFnc could also detect the essential one. It could be also interesting to determine a threshold for the VIFnc from which the multicollinearity will be considered worrying. \\
On the other hand, given the model (\ref{model0}), the expression obtained for the variance of the estimator is given by:
\begin{equation}
    \label{vari.beta.centrada}
    var (\widehat{\beta}_{j}) = \frac{\sigma^{2}}{RSS_{j}}, \quad j=1,\dots,k,
\end{equation}
where $RSS_{j}$ is the residual sum of squares (RSS) of the auxiliary regression of the $-j$ independent variable  as a function of the rest of independent variables. Taking into account the decomposition of the squared sums, this expression is equivalent to:
\begin{equation}
    \label{vari.beta.nocentrada}
    var (\widehat{\beta}_{j}) = \frac{\sigma^{2}}{TSS_{j}} \cdot VIF(j), \quad j=2,\dots,k,
\end{equation}
where $SST_{j}$ is the residual sum of squares (TSS) of the previous auxiliary regression. But, this decomposition is only verified if there is intercept in the auxiliary regression, in other case, it is not possible to state that expressions (\ref{vari.beta.centrada}) and (\ref{vari.beta.nocentrada}) are equivalent.

In this case, while the participation in the calculation of  $var (\widehat{\beta}_{j})$ was not showed, it is not appropriate to consider that the VIFnc is a factor that inflate the variance. Anyway, it is evident that the VIFnc is able to detect near multicollinearity in a linear regression model.

In the following section, we answer the following questions:
\begin{enumerate}[a)]
    \item Is the VIFnc a factor which inflate the variance?
    \item What kind of multicollinearity is able to detect?
\end{enumerate}

Finally, main results are presented in Section \ref{sec:conc}.

\section{Auxiliary non-centered regressions}
\label{sec:regression_no_centered}

This section presents the calculation of the VIFnc considering that the auxiliary regression is non-centered, it is to say, it has no intercept. Firstly, it is presented how to calculate the coefficient of determination for non-centered models.

\subsection{Non-centered coefficient of determination}

Given the linear regression (\ref{model0}) with or without intercept, it is verified the following decomposition for the sum of squares:
\begin{equation}
    \label{descom.sc.no.centered}
    \sum \limits_{i=1}^{n} y_{i}^{2} = \sum \limits_{i=1}^{n} \widehat{y}_{i}^{2} + \sum \limits_{i=1}^{n} e_{i}^{2},
\end{equation}
where $\widehat{\mathbf{y}}$ represents the estimation of the dependent variable of the model fitted by ordinary least squares (OLS) and $\mathbf{e} = \mathbf{y} - \widehat{\mathbf{y}}$ the residuals obtained from that fit.  In this case, the coefficient of determination is obtained by the following expression:
\begin{equation}
    \label{R2.no.centered}
    R_{nc}^{2} = \frac{\sum \limits_{i=1}^{n} \widehat{y}_{i}^{2}}{\sum \limits_{i=1}^{n} y_{i}^{2}} = 1 - \frac{\sum \limits_{i=1}^{n} e_{i}^{2}}{\sum \limits_{i=1}^{n} y_{i}^{2}}.
\end{equation}

Comparing the decomposition of the sums of squares given by (\ref{descom.sc.no.centered}) with the traditionally applied to calculated the coefficient of determination in models with intercept as in model (\ref{model0}):
\begin{equation}
    \label{descom.sc.centered}
    \sum \limits_{i=1}^{n} (y_{i} - \overline{\mathbf{y}})^{2} = \sum \limits_{i=1}^{n} (\widehat{y}_{i} - \overline{\mathbf{y}})^{2} + \sum \limits_{i=1}^{n} e_{i}^{2},
\end{equation}
it is noted that both coincide if the dependent variable has zero mean. If the mean is different to zero, both models present the same residual sum of squares and different explained and total sum of squares. Thus, these models lead to the same value for the coefficient of determination (and, as consequence, for the VIF) only if the dependent variable presents a mean equal to zero.

\subsection{Non-centered variance inflator factor}

The VIFnc is obtained from expression:
\begin{equation}
    VIFnc(j) = \frac{1}{1 - R^{2}nc(j)}, \quad j=2,\dots,k,
    \label{def.FIVnc.1}
\end{equation}
being $R^{2}nc(j)$ the coefficient of determination, calculated by following (\ref{R2.no.centered}), of the non-centered auxiliary regression:
\begin{equation}
    \label{reg.aux}
    \mathbf{X}_{j} = \mathbf{X}_{-j} \boldsymbol{\delta} + \mathbf{w},
\end{equation}
where $\mathbf{X}_{-j}$ is equal to the matrix $\mathbf{X}$ after eliminating the variable $\mathbf{X}_{j}$, for $j = 2,\dots, k$, and it has not a vector of ones representing the intercept.

In this case:
\begin{itemize}
    \item $\sum \limits_{i=1}^{n} X_{ij}^{2} = \mathbf{X}_{j}^{t} \mathbf{X}_{j}$, and
    \item $\sum \limits_{i=1}^{n} \widehat{X}_{ij}^{2} = \widehat{\mathbf{X}}_{j}^{t} \widehat{\mathbf{X}}_{j} = \mathbf{X}_{j}^{t} \mathbf{X}_{-j} \cdot \left( \mathbf{X}_{-j}^{t} \mathbf{X}_{-j} \right)^{-1} \cdot \mathbf{X}_{-j}^{t} \mathbf{X}_{j}$ due to $\widehat{\mathbf{X}}_{j} = \mathbf{X}_{-j} \cdot \left( \mathbf{X}_{-j}^{t} \mathbf{X}_{-j} \right)^{-1} \cdot \mathbf{X}_{-j}^{t} \mathbf{X}_{j}$.
\end{itemize}

and then:
\begin{eqnarray}
    R^{2}nc(j) &=& \frac{\mathbf{X}_{j}^{t} \mathbf{X}_{-j} \cdot \left( \mathbf{X}_{-j}^{t} \mathbf{X}_{-j} \right)^{-1} \cdot \mathbf{X}_{-j}^{t} \mathbf{X}_{j}}{\mathbf{X}_{j}^{t} \mathbf{X}_{j}}, \nonumber \\
    1 - R^{2}nc(j) &=& \frac{\mathbf{X}_{j}^{t} \mathbf{X}_{j} - \mathbf{X}_{j}^{t} \mathbf{X}_{-j} \cdot \left( \mathbf{X}_{-j}^{t} \mathbf{X}_{-j} \right)^{-1} \cdot \mathbf{X}_{-j}^{t} \mathbf{X}_{j}}{\mathbf{X}_{j}^{t} \mathbf{X}_{j}}, \nonumber \\
    VIFnc(j) &=& \frac{\mathbf{X}_{j}^{t} \mathbf{X}_{j}}{\mathbf{X}_{j}^{t} \mathbf{X}_{j} - \mathbf{X}_{j}^{t} \mathbf{X}_{-j} \cdot \left( \mathbf{X}_{-j}^{t} \mathbf{X}_{-j} \right)^{-1} \cdot \mathbf{X}_{-j}^{t} \mathbf{X}_{j}}. \label{def.FIVnc.2}
\end{eqnarray}

Thus, the VIFnc coincides to the expression given by Stewart \cite{Stewart1987} for the VIF and denoted as  $k_{j}^{2}$, it is to say, $VIFnc(j) = k_{j}^{2}$.

However, recently, Salmerón et al. \cite{Salmeron2019} showed that the index presented by Stewart has been, even by the own Stewart, misleading identified as the VIF, verifying the following relation between both measures:
\begin{equation}
    \label{StewartVIF}
    k_{j}^{2} =     vif(j) + n \cdot \frac{\overline{\mathbf{X}}_{j}^{2}}{RSS_{j}}, \quad j=2,\dots,k,
\end{equation}
where $\overline{\mathbf{X}}_{j}$ is the mean of the $-j$ variable of $\mathbf{X}$.

From expression (\ref{StewartVIF}) it is shown that the VIFnc and the VIF only coincide if the associated variable has zero mean, analogously to what happens in the decomposition of the sum of squares.

\subsubsection{Does the VIFnc inflate the variance?}

From expression (\ref{vari.beta.centrada}) and considering that expression (\ref{def.FIVnc.2}) can be rewritten as:
$$VIFnc(j) = \frac{\mathbf{X}_{j}^{t} \mathbf{X}_{j}}{RSS_{j}},$$
it is possible to obtain:
\begin{equation}
    \label{vari.beta.centrada.bis}
    var (\widehat{\beta}_{j}) = \frac{\sigma^{2}}{RSS_{j}} = \frac{\sigma^{2}}{\mathbf{X}_{j}^{t} \mathbf{X}_{j}} \cdot VIFnc(j), \quad j=2,\dots,k.
\end{equation}

It is required to establish a model as a reference to conclude if the variance has been inflated (see, for example, Cook \cite{Cook1984}). Thus, if the variables in $\mathbf{X}$ are orthogonal, it is verified that $\mathbf{X}^{t} \mathbf{X} = diag( d_{1},\dots, d_{k})$ where $d_{j} = \mathbf{X}_{j}^{t} \mathbf{X}_{j}$. In this case,  $\left( \mathbf{X}^{t} \mathbf{X} \right)^{-1} = diag( 1/d_{1},\dots, 1/d_{k})$ and, consequently:
\begin{equation}
    \label{vari.beta.centrada.bis.ortogonal}
    var (\widehat{\beta}_{j,o}) = \frac{\sigma^{2}}{\mathbf{X}_{j}^{t} \mathbf{X}_{j}}, \quad j=2,\dots,k.
\end{equation}

In this case,
$$\frac{var (\widehat{\beta}_{j})}{var (\widehat{\beta}_{j,o})} = VIFnc(j), \quad j=2,\dots,k,$$
and, then, it is possible to state that the VIFnc is a factor that inflate the variance.

\subsubsection{What kind of near multicollinearity detects the VIFnc?}
Introduction section showed that the VIFnc detects the traditional definition of non-essential near multicollinearity. When the VIF is related to the index of Stewart, see expression (\ref{StewartVIF}), it is possible to conclude that the VIFnc is also able to detect the essential near multicollinearity.

However, the introduction also showed that the VIFnc does not detect the relation between the intercept and the rest of the independent variables, which is explained when the intercept is eliminated in the auxiliary regression. This fact is contradictory to the fact that the VIFnc coincides with the index of Stewart since this measure is able to detect the non essential multicollinearity (see Salmerón et al. \cite{Salmeron2019}).

Although, the VIFnc \textit{could be fooled} including the constant as an independent variable in a model without intercept, it is to say:
$$\mathbf{y} = \beta_{1} \cdot \mathbf{X}_{1} + \beta_{2} \cdot \mathbf{X}_{2} + \beta_{3} \cdot \mathbf{X}_{3} + \mathbf{u}.$$

The following results are obtained for the VIFnc of auxiliary regressions of this model from expression (\ref{def.FIVnc.1}) for $\mathbf{X}_{1}$, $\mathbf{X}_{2}$ and $\mathbf{X}_{3}$
$$400031.4, 199921.7, 200158.3,$$
while, only considering $\mathbf{X}_{1}$ and $\mathbf{X}_{2}$ it is obtained 199921.7 and for $\mathbf{X}_{1}$ and $\mathbf{X}_{3}$ is equal to 200158.3.

Thus, considering the centered model and calculating the coefficient of determination of the auxiliary regressions as if the model was non-centered, it is possible to detect the non-essential multicollinearity.

\section{Conclusions}
\label{sec:conc}

This work analyzes the detection of near multicollinearity from non-centered auxiliary regressions obtaining that the VIF obtained from them, VIFnc, coincides with the index of Stewart. It is also provided a definition of the non-essential multicollinearity that generalizes the definition given by Marquardt and Snee \cite{MarquardtSnee}. Finally, it will be interesting as future work to determine thresholds for the VIFnc from which determine that the degree of near multicollinearity detected is worrying.

\end{document}